\documentclass[a4paper]{jpconf}

\usepackage{citesort}

\usepackage{graphicx}
\usepackage{amsmath}
\usepackage{amssymb}
\usepackage{bbm}
\usepackage{xspace}
\usepackage{color}
\usepackage{footnote}
\usepackage{comment}
\usepackage{subfigure}

\newcommand{\fig}[1]{Fig.\thinspace{}\ref{#1}}

\newcommand{\eq}[1]{Eq.\thinspace{}(\ref{#1})}

\newcommand{\eqs}[1]{Eqs.\thinspace{}(\ref{#1})}

\newcommand{\se}{Sec.\@\xspace}

\newcommand{\tcite}[1]{Ref.~\cite{#1}}

\newcommand{\Dph}{\underline{\Delta}_\mathrm{ph}(\omega)}
\newcommand{\Daux}{\underline{\Delta}_\mathrm{aux}(\omega)}
\newcommand{\jV}{$j(\phi)$ }
\newcommand{\djV}{$\partial j / \partial \phi$ }

\newcommand{\djVns}{$\partial j / \partial \phi$}

\newcommand{\nag}{{\phantom{\dagger}}}

\begin{document}

\title{Quasiparticle excitations in steady state transport across a correlated layer}

\author{Antonius Dorda, Irakli Titvinidze, and Enrico Arrigoni}

\address{Institute of Theoretical and Computational Physics, Graz University of Technology, 8010 Graz, Austria}

\ead{dorda@tugraz.at,$\,$arrigoni@tugraz.at}

\begin{abstract}
In this work we investigate the spectral and transport properties of a single correlated layer attached to two metallic leads, with particular focus on the low-energy physics. A steady state current is driven across the layer by applying a bias voltage between the leads. Extending previous work we introduce a nonzero temperature in the leads, which enables us to study the influence of quasiparticle excitations on the transport characteristics in detail. Even though the system is clearly three dimensional we obtain current-voltage curves that closely resemble those of single quantum dots. Furthermore, a splitting of the quasiparticle excitation with bias voltage is observed in the spectral function. 
\end{abstract}

\section{Introduction}

Correlated systems out of equilibrium and in particular electronic transport through quantum dots~\cite{go.sh.98,cr.oo.98,kr.sh.12,zh.ka.13} and correlated heterostructures~\cite{an.ga.99,is.og.01,oh.mu.02,oh.hw.04,ga.ah.02,zh.wa.12} have recently attracted increasing interest. Related model systems of paramount importance are the single impurity Anderson model (SIAM)~\cite{ande.61} and the Hubbard model~\cite{hubb.63,gutz.63,kana.63}. At the present time the equilibrium properties of these systems are understood to large extent~\cite{hews.93,bu.co.08,le.an.15u,voll.12,geor.04}. The so-called dynamical mean field theory (DMFT)~\cite{ge.ko.96,me.vo.89,voll.12,geor.04} was a key step in the theoretical description and understanding of Hubbard-like models and furthermore, established a link between correlated lattice systems which exhibit a Mott transition and the Kondo physics of a SIAM. Within this framework, the coherent quasiparticle excitations are described as a self-consistent Kondo effect~\cite{geor.04}. The close relation between the SIAM and the Hubbard model poses the question whether analogous behavior is seen in the transport characteristics of the two systems. Exactly this question is the topic of the investigation presented here. 

At the heart of DMFT lies the self-consistent solution of a quantum impurity model, the SIAM. An accurate description of the nonequilibrium physics of the SIAM and the related Kondo model is challenging by itself and currently intensively studied with different methods. To mention just a few, central aspects could be established with the noncrossing approximation~\cite{wi.me.94,le.sc.01,ro.kr.01}, perturbative renormalization group (RG)~\cite{ro.pa.03,sh.ro.06}, flow equations~\cite{kehr.05,fr.ke.10}, real-time RG~\cite{pl.sc.12,re.pl.14}, time-dependent density matrix RG~\cite{he.fe.09,ho.mc.09,nu.ga.13,nu.ga.15}, numerical RG~\cite{ande.08,sc.an.11} and Monte Carlo methods~\cite{we.ok.10,ha.he.07,co.gu.14}. A method recently introduced by some of us, which is well-suited for an application within nonequilibrium DMFT, is the so-called auxiliary master equation approach (AMEA)~\cite{ar.kn.13,do.nu.14}. AMEA has proven to be an accurate method for the study of the nonequilibrium steady state physics of the SIAM~\cite{do.nu.14,do.ga.15}. Special emphasis was laid on investigating the evolution of the Kondo resonance upon driving the impurity model out of equilibrium by applying a bias voltage, and we found a transition from a single peak structure to a linear splitting of the Kondo resonance with bias. Furthermore, the current-voltage characteristics obtained for different temperatures showed clear signatures of the Kondo effect~\cite{do.ga.15}.  

In the last years, rapid progress was made in the treatment of correlated lattice models out of equilibrium within DMFT, in explicitly time-dependent~\cite{fr.tu.06,ec.ko.09,ao.ts.14,ba.wo.15} and periodic or steady state situations~\cite{sc.mo.02u,jo.fr.08,okam.08,ar.ko.12}. In the study presented here we consider the special case of transport through a correlated heterostructure, consisting of a single correlated layer attached to two metallic leads at different chemical potentials. A similar setup was already treated in earlier studies with AMEA~\cite{ar.kn.13,ti.do.15u}, however, 
the influence of temperature was not investigated. Here we consider the transport and spectral properties of the system starting from a lowest temperature, which can still be well-resolved with the employed impurity solver and results in a strong quasiparticle excitation, and successively extending to larger values of $T$ up to the quasigapped regime, analogous to a Mott insulator. Besides the bias-dependent spectral function, the experimentally well-accessible current-voltage characteristics is presented.

The work is organized as follows: In \se\ref{sec:model} the investigated model is defined, in \se\ref{sec:method} we briefly introduce the nonequilibrium DMFT approach together with AMEA, and in \se\ref{sec:results} the obtained results are presented and discussed. Concluding remarks are given in \se\ref{sec:conclusio}.

\section{Model}\label{sec:model}

\begin{figure}
\begin{center}
\includegraphics[width=0.6\textwidth]{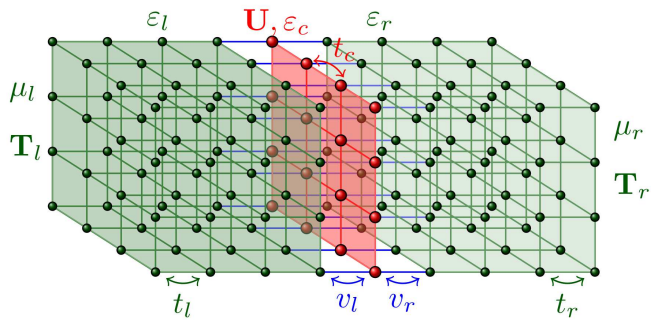}
\caption{Sketch of the investigated heterostructure, consisting of a single correlated layer of infinite size (red) with local Hubbard interaction $U$, on-site energy $\varepsilon_c$ and in-plane hopping amplitude $t_c$, sandwiched between two semi-infinite metallic leads (green). The hopping in the leads $l$ and $r$ is isotropic with amplitudes $t_l$ and $t_r$, respectively, with $v_l$ and $v_r$ denoting the  hybridizations between the respective lead and the layer. An applied bias voltage $\phi$ shifts
the on-site energies $\varepsilon_l$ and $\varepsilon_r$, as well as the chemical potentials $\mu_l$ and $\mu_r$ anti-symmetrically, which ensures together with $\varepsilon_c=-U/2$ particle-hole symmetry. All hoppings are for nearest neighbor terms only, and the temperature of the leads is labeled by $T = T_l = T_r$. We take $t_c$ as unit of energy and consider the case with $U = 10\,t_c$, $v_l = v_r = t_c$ and $t_l = t_r = 2\,t_c$. We discuss results for different values of $T$ and $\phi$. 
}
\label{fig:model}
\end{center}
\end{figure}

The model system considered in this work is schematically depicted in \fig{fig:model}. The corresponding Hamiltonian is given by
\begin{equation}
\label{eq:H}
{H}={H}_c + \sum_{\alpha=l,r}{H}_\alpha +{H}_{\rm coup} \, ,
\end{equation}
consisting of a part for the central system ${H}_c$, a part for each decoupled lead ${H}_{l/r}$ and a coupling between the leads and the correlated layer ${H}_{\rm coup}$. In the detail the Hamiltonian reads 
\begin{eqnarray}
{H}_c&=& -t_c \sum_{\langle ij\rangle, \sigma}c_{i\sigma}^\dagger c_{j\sigma}^{\phantom\dagger} +U\sum_i n_{i\uparrow}n_{i_\downarrow} +\varepsilon_c\sum_{i,\sigma}n_{i\sigma} \,,\\
{H}_\alpha&=& -t_\alpha \sum_{\langle i j\rangle  \sigma}c_{\alpha i \sigma }^\dagger c_{\alpha j \sigma }^{\phantom\dagger} 
+\varepsilon_\alpha\sum_{i \sigma}c_{\alpha i\sigma}^\dagger c_{\alpha i \sigma }^{\phantom\dagger} \,,\\
{H}_{\rm coup}&=&\sum_{\langle i j\rangle  \alpha \sigma} v_\alpha\left(c_{i\sigma}^\dagger c_{ \alpha j\sigma}^{\phantom\dagger} + h.c.\right) \, ,
\end{eqnarray}
where $ \langle i,j\rangle$ indicates neighboring sites, $c_{i,\sigma}^\dagger$ creates an electron at the $i$-th site of the correlated layer with spin $\sigma=\uparrow,\downarrow$, and $n_{i\sigma}=c_{i \sigma}^\dagger c_{i \sigma}^{\phantom\dagger}$ denote the corresponding occupation number operators. The analogous fermionic creation/annihilation operators of lead $\alpha$ are labeled by $c_{ \alpha i \sigma}^\dagger/c_{ \alpha i \sigma}^\nag$. For the particular parameters see \fig{fig:model}.

\section{Method}\label{sec:method}

In the following we outline the method only briefly and for details we refer to \tcite{ar.kn.13,do.nu.14,ti.do.15u}.

Nonequilibrium dynamics are conveniently formulated in terms of Keldysh Green's functions~\cite{kad.baym,schw.61,keld.65,ha.ja,ra.sm.86,le.da.06}, whereby for the steady state limit it suffices to consider $2\times2$ objects on the Keldysh contour
\begin{equation}
\underline G =\left(
\begin{array}{cc}
G^R & G^K \\
0   & G^A 
\end{array}
\right) \,,
\label{eq:KeldyshGF}
\end{equation}
which we denote by an underscore. Here, the retarded $G^R$ and the Keldysh component $G^K$ are independent functions in a generic nonequilibrium situation, and the advanced part is given by $G^A=(G^R)^\dagger$. The spectral function is defined as usual: $A = i/2\pi\,(G^R-G^A)$. 

Since the model outlined in \se\ref{sec:model} is translationally invariant in the in-plane direction, it is convenient to introduce the corresponding momentum variable ${\bf k}_{||}$. Furthermore, time translational invariance applies in the steady state limit and the governing equations can be formulated in the frequency domain $\omega$. With this the Green's function of the correlated layer is given in terms of Dyson's equation by
\begin{equation}
{\underline G}^{-1}(\omega, {\bf k}_{||}) ={\underline g}^{-1}_0(\omega, {\bf k}_{||}) - \sum_{\alpha=l,r}v_\alpha^2~{\underline g}_\alpha(\omega, {\bf k}_{||}) - {\underline{\Sigma}}(\omega, {\bf k}_{||}) \,.
\label{eq:Dyson}
\end{equation}
Here, the decoupled non-interacting Green's function of the layer is denoted by ${\underline g}_0(\omega, {\bf k}_{||})$, and those of the leads by ${\underline g}_\alpha(\omega, {\bf k}_{||})$. The non-interacting Green's functions are known exactly but the determination of the self-energy of the correlated layer ${\underline{\Sigma}}(\omega, {\bf k}_{||})$ is challenging and one has to resort to approximations. In particular we neglect spatial correlations and restrict ourselves to a local self-energy ${\underline{\Sigma}}(\omega, {\bf k}_{||}) ={\underline{\Sigma}}(\omega)$, as usually done in the context of DMFT~\cite{ge.ko.96,me.vo.89,geor.04,voll.12,sc.mo.02u,fr.tu.06}. Within DMFT, the local quantity ${\underline{\Sigma}}(\omega)$ is determined by mapping the lattice problem onto an equivalent quantum impurity model, with the same local parameters $U$ and $\varepsilon_c$. However, the bath degrees of freedom of the impurity model depend on ${\underline{\Sigma}}(\omega)$, such that a self-consistent solution is needed, which is commonly obtained in an iterative manner.
The bath for the impurity model is fully specified by the hybridization function
\begin{equation}
\underline{\Delta}_\mathrm{ph}(\omega)=\underline{g}_0^{-1}(\omega) - \underline{G}^{-1}_{\rm loc}(\omega)-\underline{\Sigma}(\omega)\,, 
\label{eq:D_ph}
\end{equation}
where $\underline{g}_0^{-1}(\omega)$ is the non-interacting Green's function of the disconnected impurity and the local Green's function is obtained by
\begin{equation}
\underline{G}_{\rm loc}(\omega)=\int\limits_{\rm BZ} \frac{d{\bf k}_{||}}{(2\pi)^2}\underline{G}(\omega,{\bf k}_{||}) \,.
\label{eq:G_loc}
\end{equation}
In order to solve the nonequilibrium impurity problem we resort to AMEA, cf. \cite{ar.kn.13,do.nu.14}, which maps the original impurity problem onto an auxiliary one, with a finite number of bath sites $N_B$ and additional Markovian environments. The resulting open quantum system is described by a Lindblad equation and is small enough to be solved accurately by numerical techniques. In contrast to other approaches, the parameters of the Lindblad equation are not obtained perturbatively but through an optimization procedure. In particular, we consider the hybridization function of the auxiliary system $\Daux$ and vary the auxiliary bath parameters in order to minimize the difference to the physical hybridization function $\Dph$, \eq{eq:D_ph}. In the limit of large $N_B$ the approach becomes exact but even for small values of $N_B$ we obtain $\Daux \approx \Dph$ to very good approximation. Typically, an exponential convergence with increasing $N_B$ is achieved. After the mapping procedure, which can be done in a $U=0$ calculation, the interacting impurity problem is solved. For this we introduced two different strategies in previous work: On the one hand, an implementation of AMEA which makes use of Krylov space methods, cf.~\tcite{do.nu.14}, and on the other hand, a matrix product states based solution, cf.~\tcite{do.ga.15}. The latter allows for a highly accurate solution of the impurity problem but requires a rather large amount of CPU time. The former is not as accurate at low temperatures but faster in many cases and is used in the present work. The Krylov space solver enables us to consider up to $N_B=6$ bath sites, which suffices to treat cases with strong Kondo or quasiparticle excitations reliably, cf.~\tcite{do.nu.14,ti.do.15u}. Once the many-body problem of the auxiliary system is solved one obtains an approximation for the self-energy
\begin{equation}
\underline{\Sigma}(\omega)=\underline{G}_{{\rm aux},0}^{-1}(\omega)-\underline{G}_{\rm aux}^{-1}(\omega) \,,
\label{eq:Sigma}
\end{equation}
from the knowledge of the non-interacting and the interacting auxiliary Green's functions. By inserting $\underline{\Sigma}(\omega)$ from \eq{eq:Sigma} into \eqs{eq:Dyson},(\ref{eq:D_ph}),(\ref{eq:G_loc}) we close the DMFT cycle and iterate until a self-consistent point is reached.

\section{Results}\label{sec:results}

\begin{figure*}
\begin{center}
\includegraphics[width=0.47\textwidth]{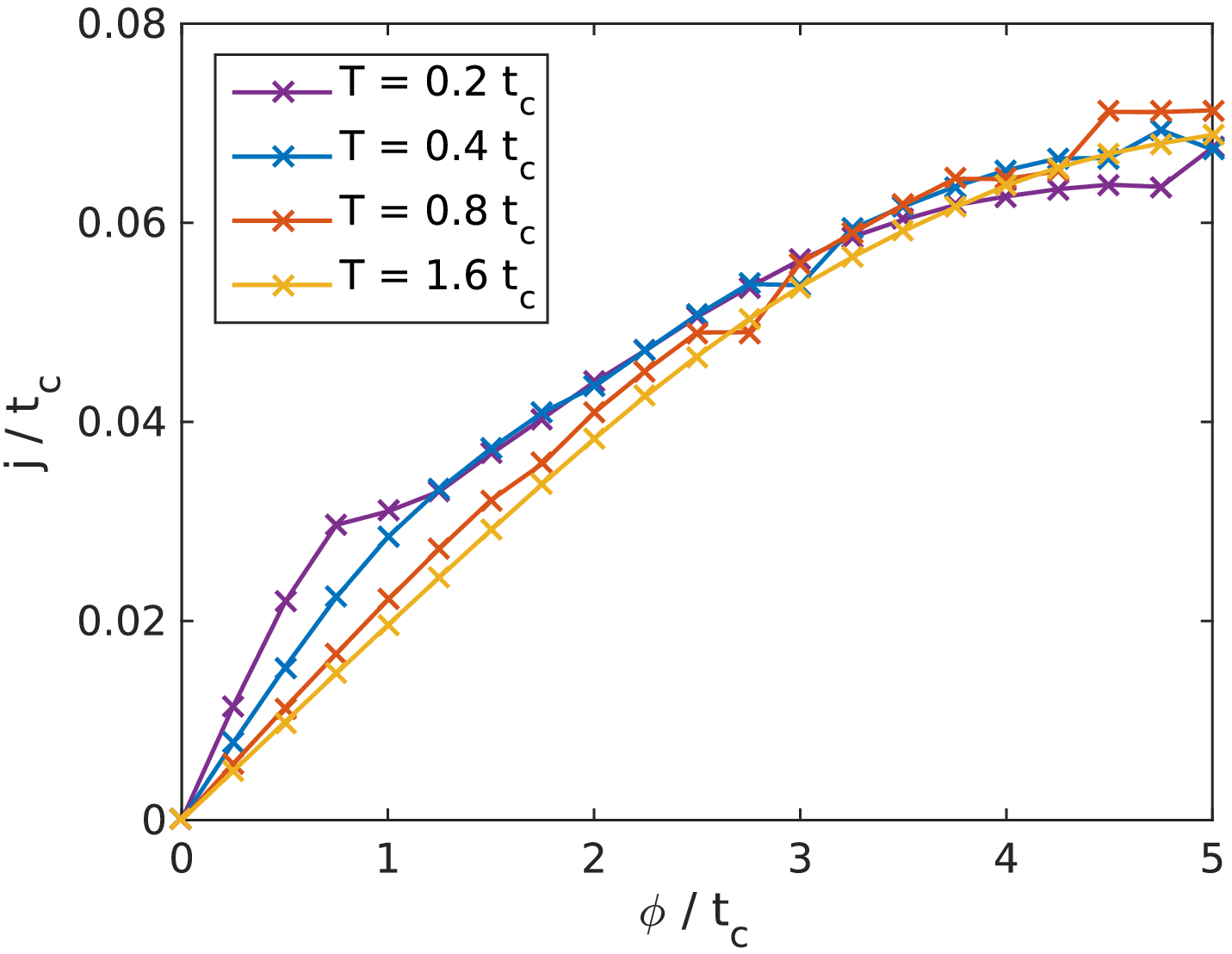}
\includegraphics[width=0.47\textwidth]{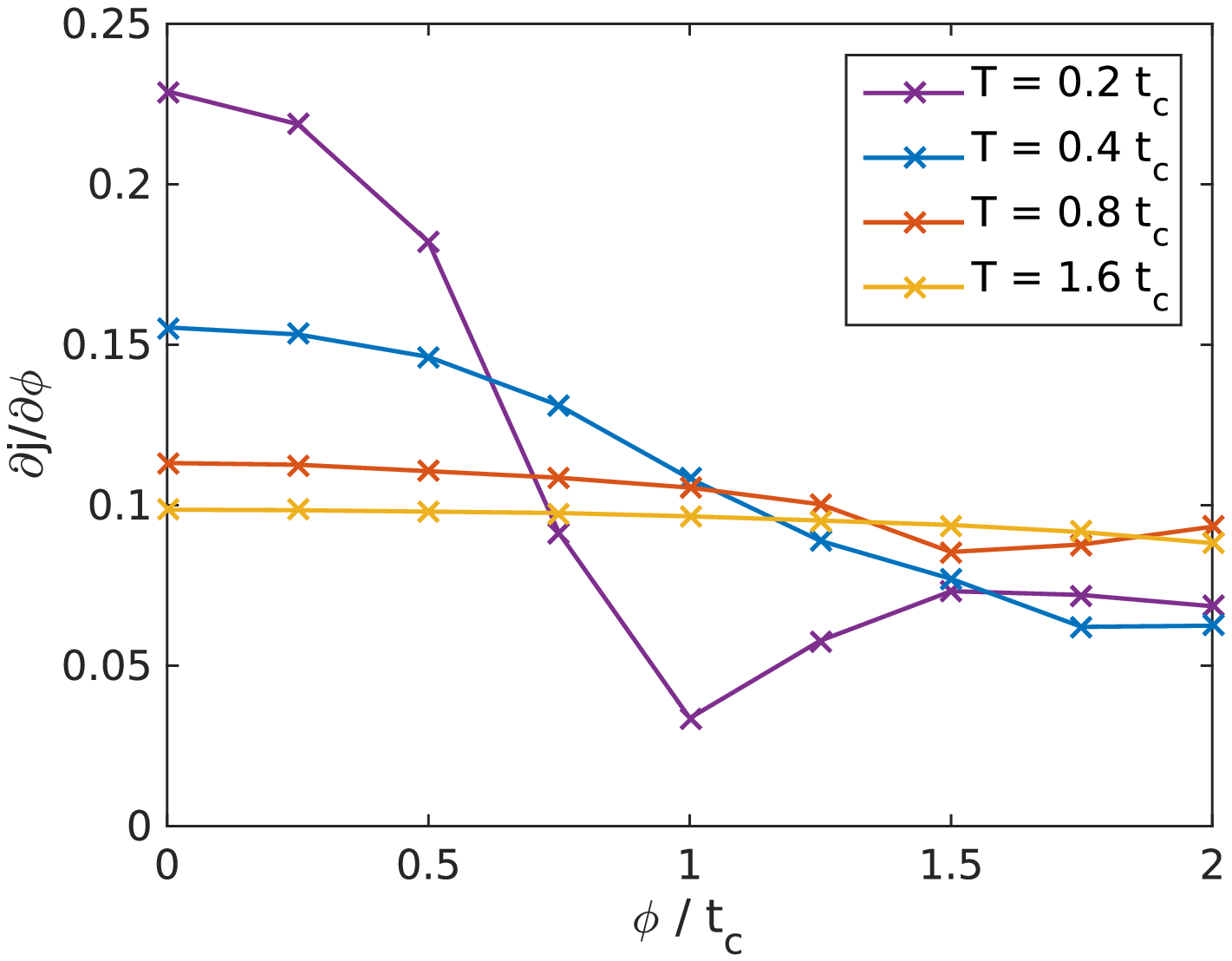}
\caption{Current \jV and differential conductance \djV as a function of bias voltage $\phi$ and for different temperatures $T$. For the particular model parameters see \fig{fig:model}. Calculations for $T = 1.6\,t_c$ were performed with $N_B = 4$ and all others with $N_B = 6$.}
\label{fig:current}
\end{center}
\end{figure*}
\begin{figure*}
\begin{center}
\includegraphics[width=0.95\textwidth]{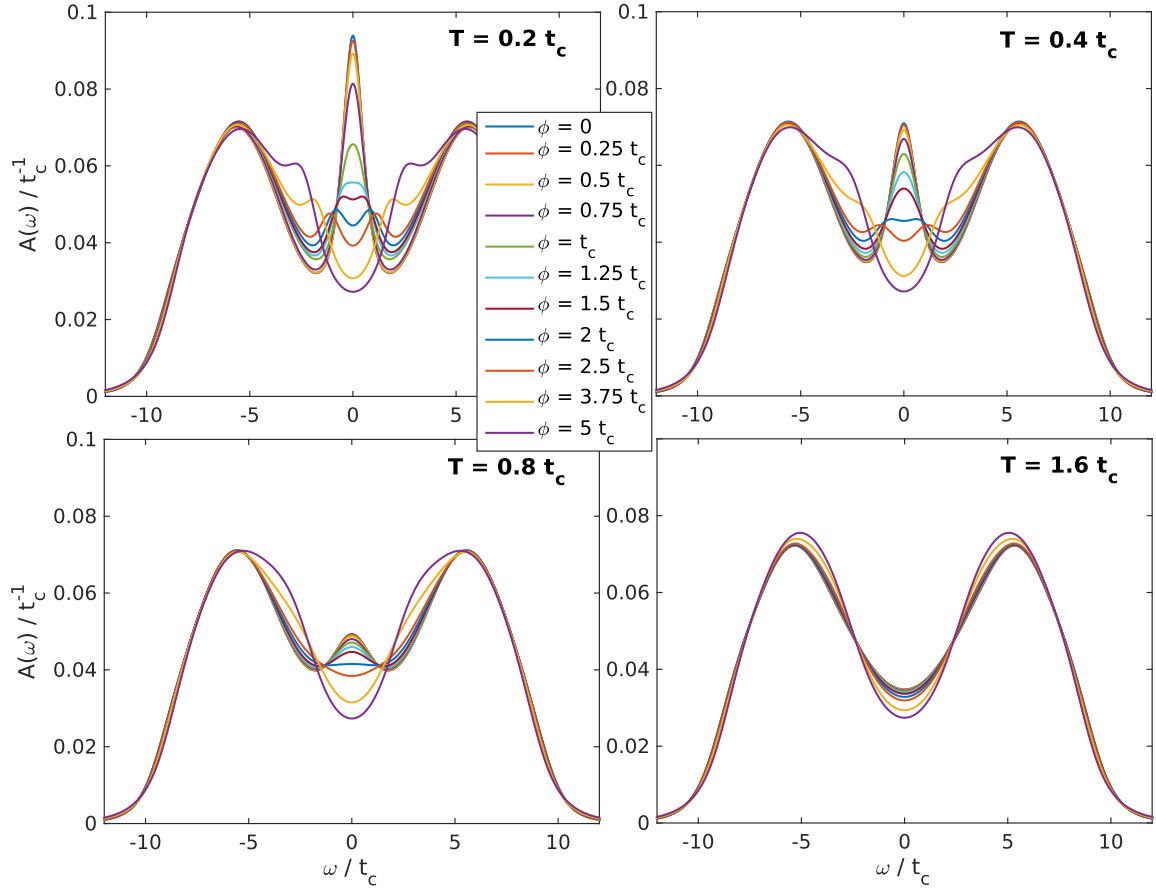}
\caption{Spectral function $A(\omega)$ as a function of bias voltage $\phi$ and for different temperatures $T$. 
The parameters are the same as in \fig{fig:current}}
\label{fig:spectral_function}
\end{center}
\end{figure*}
\begin{figure*}
\begin{center}
\includegraphics[width=0.95\textwidth]{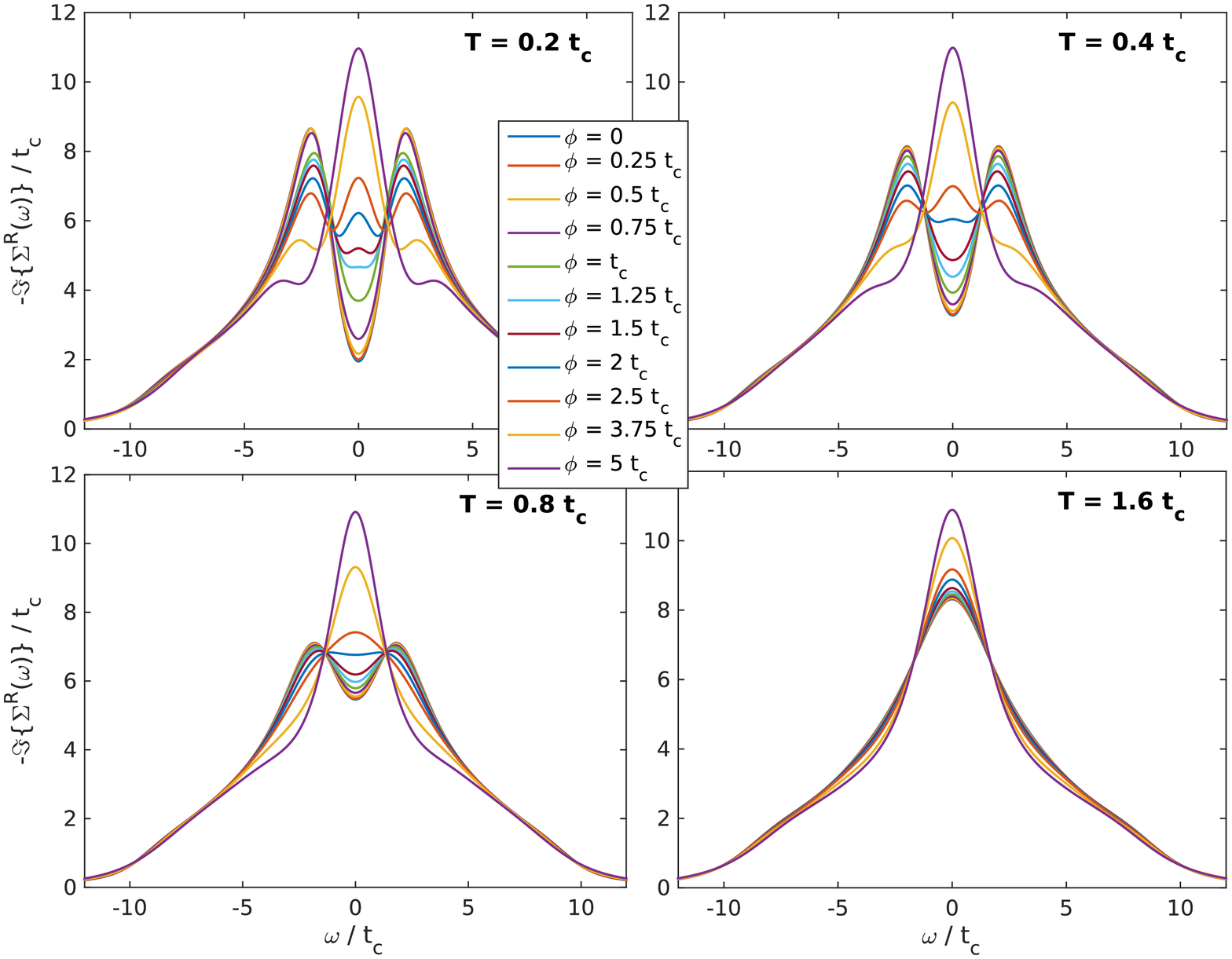}
\caption{Retarded self-energy $\Sigma^R(\omega)$ as a function of bias voltage $\phi$ and for different temperatures $T$.
The parameters are the same as in \fig{fig:current}}
\label{fig:self_energy}
\end{center}
\end{figure*}

In the following we present results for the transport and spectral properties of a correlated layer in a nonequilibrium steady state situation. The particular model is defined in \se\ref{sec:model} and depicted in \fig{fig:model}. We place special emphasis on the low-energy properties and consider cases with rather low bias voltages $\phi$. To investigate the role of resonant quasiparticle excitations, different temperatures $T = T_l = T_r$ are introduced in the leads. In contrast to previous work \cite{ti.do.15u}, a nonzero temperature is considered in the leads and additionally, the hopping parameters are chosen such that each correlated site has an equal hopping amplitude to neighboring sites ($v_l = v_r = t_c$). By this we expect to be in a regime in which a competition between the physics of an isolated 2D layer and the one of single quantum dots occurs. 

In \fig{fig:current} the current-voltage characteristics \jV together with the differential conductance \djV are displayed for various $T$. At low bias voltages $\phi \lesssim 2\,t_c$ the temperature has large influence on \jV and \djVns, whereas for larger voltages $\phi \gtrsim 2.5\,t_c$ we find quantitatively similar current values for all of the temperatures. This is in close analogy to what was observed in \tcite{do.ga.15} for the nonequilibrium properties of a quantum dot. In \djV we obtain for $\phi \lesssim 1\,t_c$ a strong dependence on $T$. Again, the behavior is similar to what is known from Kondo systems~\cite{kr.sh.12,zh.ka.13,pl.sc.12,re.pl.14}. However, the accuracy of the present calculations does not permit us to investigate the scaling with $\phi$ in detail, in particular if a logarithmic dependence as in quantum dots is present. Still, one observes that the experimentally well-accessible quantity \djV exhibits a strong temperature dependence and clear signatures of a Kondo-like behavior are visible for low $T$. From a technical point of view one should note that slight kinks or jumps in \jV are present, at different values of $\phi$ for the various $T$. These artefacts originate in the mapping procedure and appear at values $\phi = \phi_c$ at which more than one parameter set for the auxiliary system minimizes the difference to the physical hybridization function. Usually, one of these minima is better for $\phi<\phi_c$ while the other one for $\phi>\phi_c$. Therefore, such a crossing of minima leads to an abrupt change in parameters of the auxiliary system and may result in a slight shift of spectral weight, cf. \tcite{do.nu.14}. In general, this effect is smallest in situations in which the difference between $\Daux$ and $\Dph$ is small in any case and thus, is reduced upon increasing the number of bath sites $N_B$.

A more detailed picture of the state of the system is obtained by investigating the spectral function, see \fig{fig:spectral_function}. In the equilibrium case, $\phi=0$, we find a strong quasiparticle excitation at $\omega=0$ for $T = 0.2\,t_c$, which is attenuated with increasing temperature ($T = 0.4\,t_c$ and $T = 0.8\,t_c$) and is completely suppressed for $T = 1.6\,t_c$.\cite{footnote_friedel} Especially interesting are the low temperature situations $T = 0.2\,t_c$ and $T = 0.4\,t_c$ in which a strong zero frequency excitation is visible in equilibrium, and for which we observe with increasing $\phi$ at first a reduction of the peak height before a splitting sets in. Similar to a quantum dot system \cite{do.nu.14,do.ga.15}, we find two resonant excitations at the Fermi-energies of the two leads and thus a linear splitting with $\phi$. For the case of $T = 0.2\,t_c$ the excitations are still clearly visible at rather large voltages up to $\phi \approx 5\,t_c$. The results for $T = 0.8\,t_c$ and $T = 1.6\,t_c$ reveal dissimilar spectral functions at low bias voltages, but surprisingly, the obtained current values in \fig{fig:current} are comparable. The reason for this is that the high temperatures average out details in $A(\omega)$ to large extent. 

The presence of resonant excitations is also clearly visible in the retarded part of the self-energy, displayed in \fig{fig:self_energy}. For $\phi=0$ we find for temperatures up to $T = 0.8\,t_c$ a local minimum in $-\Im\{\Sigma^R(\omega)\}$ at $\omega=0$, indicating a quasiparticle excitation. But only in the cases $T = 0.2\,t_c$ and $T = 0.4\,t_c$ the temperature induced decoherence is weak enough to obtain a splitting of the single minimum when increasing the bias voltage. In contrast, the self-energy for $T = 1.6\,t_c$ is rather featureless and only weakly dependent on the bias voltage.

\section{Conclusions}\label{sec:conclusio}

In this work we investigated the steady state properties of a correlated layer sandwiched between two metallic leads at different chemical potentials, induced by an externally applied bias voltage. For this we made use of a nonequilibrium DMFT approach together with AMEA as impurity solver. In addition to previous work~\cite{ti.do.15u}, we studied the influence of temperature on the transport characteristics and on the bias-dependent spectral function, with focus on the low-bias regime. The parameters of the system were chosen such that a certain direction was not preferred in advance. In particular, all of the hopping amplitudes of a correlated site to its neighbors were of equal size. From investigating the spectral function and the differential conductance as a function of bias voltage and for various temperatures, we found that the considered system bore close analogy to the case of a single quantum dot. A result like this could be expected when considering the limit in which the hopping parallel to the 2D layer is much smaller than the longitudinal one, regarding the layer and the leads. But, since the hoppings to correlated sites were isotropic the result is not intuitive nor trivial.

\section*{Acknowledgements}\label{ac}

We acknowledge valuable discussions with Martin Nuss, Markus Aichhorn, Michael Knap, and Wolfgang von der Linden. 
This work was supported by the Austrian Science Fund (FWF): P24081, P26508, as well as SFB-ViCoM project F04103, and NaWi Graz. The calculations were partly performed on the D-cluster Graz and on the VSC-3 cluster Vienna.

\section*{References}
\bibliographystyle{iopart-num}
\bibliography{ndmft_QP.bbl}

\begin{thebibliography}{10}
\bibitem{footnote_friedel}
One should note that the $T=0$ Fermi liquid limit is given by
  $A(\omega=0)\approx0.1667\,t_c^{-1}$. The accurate consideration of
  temperatures below $T = 0.2\,t_c$ would require more bath sites $N_B$ and
  thus more advanced manybody solution strategies, cf. \tcite{do.ga.15}.
\end{thebibliography}

\end{document}